\documentclass[prl,aps,twocolumn]{revtex4}
\usepackage{graphicx}

\begin{document}

\preprint{PCP-19/ISS2002}

\title{Resonant two-magnon Raman scattering in two-dimensional and ladder-type Mott insulators}

\author{Hiroaki Onodera, Takami Tohyama and Sadamichi Maekawa}
\affiliation{Institute for Materials Research, Tohoku University, Sendai, 980-8577, Japan.}
\date{\today}

\begin{abstract}
We investigate the resonant two-magnon Raman scattering in the two-dimensional (2D) and ladder-type Mott insulators by using a half-filled Hubbard model in the strong coupling limit. By performing numerical diagonalization calculations for small clusters, we find that the model can reproduce the experimental features in the 2D that the Raman intensity is enhanced when the incoming photon energy is not near the absorption edge but well above it. In the ladder-type Mott insulators, the Raman intensity is found to resonate with absorption spectrum in contrast to the 2D system. The difference between 2D and the ladder systems is explained by taking into account the fact that the ground state in 2D is a spin-ordered state while that in ladder is a spin-gapped one.

%
\end{abstract}


\maketitle

In Mott insulators, the nature of the photoexcited states is less clarified as compared with that of low-energy spin states that are described by the Heisenberg model. The excitations to the photoexcited states are observed by measurements of optical absorption. In addition, resonant Raman scattering is very useful for the investigation of the photoexcited states, because the resonance occurs when incoming photon energy $\omega_\mathrm{i}$ is near the energy of the photoexcited states.

In two-dimensional (2D) insulating cuprates that are typical Mott insulators, a two-magnon peak is observed in the Raman scattering~\cite{lyons}. Two-magnon is a magnetic excitation from the ground state of the Mott insulators. The $\omega_\mathrm{i}$ dependence of the two-magnon Raman intensity in the B$_\mathrm{1g}$ geometry shows an interesting feature~\cite{yoshida,blumberg}: The two-magnon Raman intensity is not enhanced at near the absorption edge ($\sim1.8$-$2$~eV), but a resonance occurs at higher $\omega_\mathrm{i}$ ($\sim3$~eV).

Two-magnon peaks are also observed in the ladder-type Mott insulators~\cite{sugai}. In contrast to the 2D system, the two-magnon Raman intensity in the ladder compound Sr$_{14}$Cu$_{24}$O$_{41}$ resonates with absorption spectrum~\cite{gozar}, i.e., the intensity is enhanced at the absorption edge ($\sim 2$~eV).

In this paper, we theoretically investigate the $\omega_\mathrm{i}$ dependence of the two-magnon Raman intensity in the 2D and ladder-type Mott insulators, and we clarify the origin of the difference of the $\omega_\mathrm{i}$ dependence between the 2D and ladder systems. There have been several theoretical studies about the two-magnon Raman scattering in the ladder system, where only the non-resonant regions were examined~\cite{natsume,freitas,schmidt}. In contrast to them, our study deals with resonant regions. We find that numerical calculations for the Hubbard model in the strong coupling limit reproduce the difference of the $\omega_\mathrm{i}$ dependence of the two-magnon Raman intensity at the absorption edge between the 2D and ladder-type Mott insulators. The difference is caused by the difference of spin states in the two insulators.

The Hamiltonian of the Hubbard model reads
\begin{equation}
H=-t \sum_{\langle i,j \rangle, \sigma}(c_{i,\sigma}^{\dagger} c_{j,\sigma}+H.c.)
+U\sum_{i}n_{i,\uparrow} n_{i,\downarrow},
\label{hubb}
\end{equation}
where the summations $\langle i,j \rangle$ run over nearest-neighbor pairs. We consider the strong coupling limit, i.e., $U\gg t$. In this limit, the ground state $\left| i \right\rangle$ and the final states of the two-magnon Raman scattering $\left| f \right\rangle$ have no doubly occupied site, and they are described by the Heisenberg model with nearest-neighbor exchange interaction $J(=4t^2/U)$. On the other hand, in the intermediate states $\left| \mu \right\rangle$ of the Raman process, there are one doubly occupied site and one vacant site. In this case, an effective Hamiltonian of (\ref{hubb}) is given by
\begin{eqnarray}
H_\mathrm{eff}&=&\Pi_1 H_t \Pi_1-U^{-1} \Pi_1 H_t \Pi_2 H_t \Pi_1 \nonumber \\
&+&U^{-1} \Pi_0 H_t \Pi_1 H_t \Pi_0+U,
\label{Heff}
\end{eqnarray}
where $H_t$ is the hopping term of $H$, and $\Pi_0$, $\Pi_1$, and $\Pi_2$ are projection operators onto the Hilbert space with zero, one, two doubly occupied sites, respectively. A complete expression of Eq.~(\ref{Heff}) has been given elsewhere~\cite{takahashi}. The value of $U$ is evaluated to be $U$=10$t$ for the gap values to be consistent with experimental ones.

The Raman scattering intensity at the Raman shift energy $\omega$ is expressed as
$R(\omega)=\sum_f \left| \langle f \left| M_R \right| i \rangle \right|^2 \delta \left( \omega -E_f + E_i \right)$,
where $E_i$ and $E_f$ are the ground-state and final-state energies, respectively. $M_R$ is the Raman tensor operator whose matrix element is given by
\begin{equation}
\langle i | M_{\mathrm{R}}^{\alpha,\beta} | f \rangle = \sum_{\mu}
\frac{\langle f | j_{\alpha} | \mu \rangle \langle \mu | j_{\beta} | i \rangle}
{E_{\mu}-E_{i}-\omega_{i}+i \Gamma},
\end{equation}
where $\alpha$ and $\beta$ are polarizations of the incoming and scattering photons, respectively, $j_{\alpha}$ is the current operator to the $\alpha$ direction and, $E_{\mu}$ and $\Gamma$ are energy and a damping rate for the intermediate states, respectively. Absorption spectrum $\varepsilon_2$ for the $\alpha$ polarization is given by $\omega_\mathrm{i}^{-2} \sum_{\mu} |\langle i|j_{\alpha}|\mu \rangle|^2 \delta(\omega_\mathrm{i} -E_{\mu} +E_{i})$. We calculate $R\left( \omega \right)$ and $\varepsilon_2$ by using the numerical diagonalization method for small clusters ($\sqrt{20}\times\sqrt{20}$ square and 10 $\times$ 2 ladder lattices). We use periodic boundary conditions in the 2D system. In the ladder system, a periodic boundary condition is used along the leg direction (the $x$ direction), while an open boundary condition is used along the rung direction (the $y$ direction).

\begin{figure}[t]
\includegraphics[width=7.5cm]{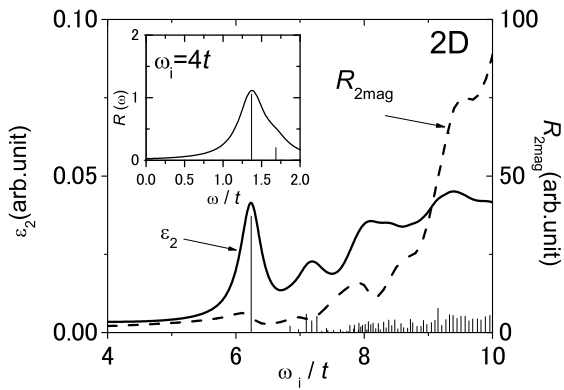}
\caption{The $\omega_\mathrm{i}$ dependence of Raman intensity $R_\mathrm{2mag}$ (dashed line), and absorption spectrum $\varepsilon_2$ (solid line) in a 20-site 2D cluster with $U=10t$. Inset shows two-magnon Raman spectrum $R\left( \omega\right)$ at $\omega_\mathrm{i}=4.0t$ . In $\varepsilon_2$ and $R\left( \omega\right)$, the solid lines are obtained by performing a Lorentzian broadening with of $0.4t$ on delta functions denoted by thin vertical bars.}
\label{fig1}
\end{figure}

First, we show results in the B$_{1g}$ geometry for the 2D system in Fig.~\ref{fig1}~\cite{tohyama}. The inset shows $R\left( \omega \right)$ at $\omega_\mathrm{i}=4.0t$. We define $R_\mathrm{2mag}$ as the integrated intensity of the peaks at $0 < \omega < 2.0t$. The $\omega_\mathrm{i}$ dependence of $R_\mathrm{2mag}$  (dashed line) and $\varepsilon_2$ (solid line) are shown in the main panel. $\varepsilon_2$ has an edge peak at $\omega_\mathrm{i}=6.2t$. However, $R_\mathrm{2mag}$ is not enhanced at the peak. This $\omega_\mathrm{i}$ dependence of $R_\mathrm{2mag}$ is similar to the experimental data~\cite{yoshida,blumberg}.

\begin{figure}
\includegraphics[width=7.5cm]{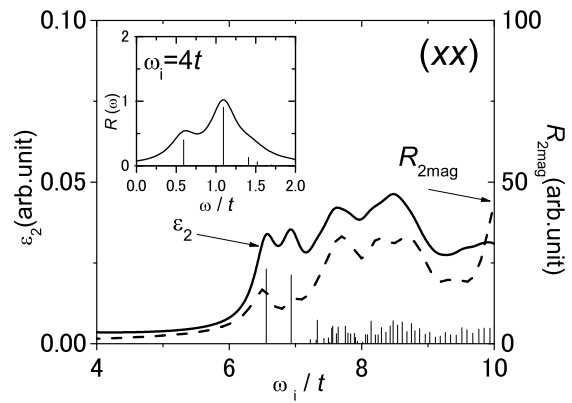}
\caption{The $\omega_\mathrm{i}$ dependence of Raman intensity $R_\mathrm{2mag}$ (dashed line), and absorption spectrum $\varepsilon_2$ (solid line) in a $10\times 2$ ladder cluster with $U=10t$ for the $(xx)$ polarization. Inset shows two-magnon Raman spectrum $R\left( \omega\right)$ at $\omega_\mathrm{i}=4.0t$. In $\varepsilon_2$ and $R\left( \omega\right)$, the solid lines are obtained by performing a Lorentzian broadening with of $0.4t$ on delta functions denoted by thin vertical bars.}
\label{fig2}
\end{figure}

Next, we discuss results for the ladder-type Mott insulators. We calculate $R\left( \omega \right)$ for the $(xx)$ polarization, where $(xx)$ means that the polarizations of both the incoming and scattering photons are along the $x$ direction. The values of hoppings (exchange interactions) are taken to be $t_\mathrm{leg}=t_\mathrm{rung}$ ($J_\mathrm{leg}=J_\mathrm{rung}$). We note that the following results are qualitatively unchanged even if we use a realistic rate, $J_\mathrm{rung}/J_\mathrm{leg}=0.8$, corresponding to Sr$_{14}$Cu$_{24}$O$_{41}$~\cite{gozar}. In Fig.\ref{fig2}, the inset shows $R\left( \omega \right)$ for the $(xx)$ polarization at $\omega_\mathrm{i}$=$4.0t$. The spectrum spreads in the range of $0.5t < \omega < 1.5t$, but the main peak appears at $\omega \sim 1.0t$. This is consistent with the fact that the peak obtained by the non-resonant Raman scattering calculations~\cite{natsume,schmidt} is located at $\omega \sim 1.0t$. The $\omega_\mathrm{i}$ dependence of $R_\mathrm{2mag}$  (dashed line) for the $(xx)$ polarization and $\varepsilon_2$ (solid line) for the $x$ polarization is shown in the main panel. $\varepsilon_2$ has an edge peak at $\omega_\mathrm{i}$=$6.5t$. $R_\mathrm{2mag}$ is enhanced at the peak in contrast to the 2D system, being consistent with experimental features.

Let us consider the origin of the different $\omega_\mathrm{i}$ dependence between the 2D and ladder systems. In the 2D Mott insulators, the ground state $\left| i\right\rangle$ is dominated by N\'eel-type spin configurations because of the presence of antiferromagnetic order, while the weight of the N\'eel-type spin configurations in the two-magnon final state $\left| f\right\rangle$ is very small~\cite{tohyama}. Since the photoexcited state at the absorption edge, $\left| \mu_\mathrm{edge} \right\rangle$, significantly contains the N\'eel-type spin configurations in the spin background, $\left\langle \mu_\mathrm{edge} \right| j_{x} \left| i \right\rangle$ becomes large but $\left\langle f \right| j_{x} \left| \mu_\mathrm{edge} \right\rangle$ is very small because of a small amount of the N\'eel-type spin configurations in $\left| f\right\rangle$. Therefore, $R_\mathrm{2mag}$, which is proportional to $\left| \left\langle f \right| j_x \left| \mu_\mathrm{edge} \right\rangle \left\langle \mu_\mathrm{edge} \right| j_x \left| i \right\rangle \right|^2$, is not enhanced at the absorption edge.  On the other hand, in the ladder system we find that the weight of the N\'eel-type spin configurations in $\left| i \right\rangle$ is smaller than that of 2D but the weight in $\left| f \right\rangle$ at $\omega=0.6t$ is as large as that in $\left| i \right\rangle$ and is thus much larger than that of 2D.  As a result, $\left\langle f \right| j_x \left| \mu_\mathrm{edge} \right\rangle$ is enhanced as compared with that in 2D and thus $R_\mathrm{2mag}$ is enhanced at the absorption edge. This is intuitively understood in the following way. The two-leg ladder has a spin gap and thus the ground state is a spin-liquid state in contrast with the 2D system where antiferromagnetic order exists.  While in 2D the two-magnon final state is very different from the initial state with respect to the N\'eel-type spin spin configurations, the difference between the initial and final states in ladder is small because both the states are of spin liquid and contain a similar amount of the N\'eel-type spin configurations. This clearly explains the different $\omega_\mathrm{i}$ dependence between 2D and ladder.

\begin{figure}[t]
\includegraphics[width=7.5cm]{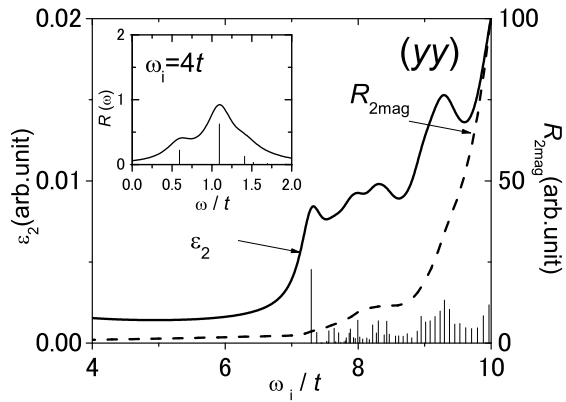}
\caption{The same as Fig.~\ref{fig3} but for the $(yy)$ polarization.}
\label{fig3}
\end{figure}

Finally, we show results of the two-magnon Raman spectrum for the $(yy)$ polarization, i.e., the poralization along the rung direction. In Fig.~\ref{fig3}, the inset shows $R\left( \omega\right)$  at $\omega_\mathrm{i}=4.0t$. The spectral shape is almost the same as that for the $(xx)$ polarization at $\omega_\mathrm{i}=4.0t$. We note that in the calculation of the non-resonance Raman scattering the same shapes are obtained for the $(xx)$ and $(yy)$ polarizations in the case of $J_\mathrm{leg}=J_\mathrm{rung}$~\cite{freitas}. The main panel in Fig.~\ref{fig3} shows the $\omega_\mathrm{i}$ dependence of $R_\mathrm{2mag}$ (dashed line) for the $(yy)$ polarization and $\varepsilon_2$ (solid line) for the $y$ polarization. $\varepsilon_2$ exhibits an edge peak at $\omega_\mathrm{i}=7.3t$. $R_\mathrm{2mag}$ is resonantly enhanced at $\omega_\mathrm{i}\sim 10t$, but not at the absorption edge in contrast to the $(xx)$ polarization. This is, however, in contradiction with the experiment where $R_\mathrm{2mag}$ for $(yy)$ is enhanced at the absorption edge~\cite{gozar}. This disagreement may come from the fact that we do not consider the interactions between ladders from the theoretical viewpoint and a small amount of holes are doped into the ladders in Sr$_{14}$Cu$_{24}$O$_{41}$ from the experimental side.

In summary, we have investigated the dependence of the two-magnon Raman intensity on the incoming photon energy $\omega_\mathrm{i}$ in the 2D and ladder-type Mott insulators by using the Hubbard model in the strong coupling limit. We have found that the difference of spin states between the 2D and ladder systems causes different $\omega_\mathrm{i}$ dependence of the two-magnon Raman intensity at the absorption edges.

 This work was supported by a Grant-in-Aid for scientific Research from the Ministry of Education, Culture, Sports, Science and Technology of Japan, and CREST. The numerical calculations were performed in the supercomputing facilities in ISSP, University of Tokyo, and IMR, Tohoku University.


\begin{thebibliography}{13}

\bibitem{lyons}
K. B. Lyons, P. A. Fleury, J. P. Remeika, A. S. Cooper, T. J. Negran, Phys.
Rev. B {\bf 37} (1988) 2353.
\bibitem{yoshida}
M. Yoshida, S. Tajima, N. Koshizuka, S. Tanaka, S. Uchida, T. Itoh,
Phys. Rev. B {\bf 46} (1992) 6505.
\bibitem{blumberg}
G. Blumberg, P. Abbamonte, M. V. Klein, W. C. Lee, D. M. Ginsberg, L. L. Miller, A. Zibold,
Phys. Rev. B {\bf 53} (1996) R11930.
\bibitem{sugai} S. Sugai, in {\it Magneto-Optics}, edited by S. Sugano, N. Kojima (Springer, Berlin, 2000), p. 75.
\bibitem{gozar} A. Gozar, G.  Blumberg, B. S. Shastray, N. Motoyama, H. Eisaki, S. Uchida, Phys. Rev. Lett. {\bf 87} (2001) 19702.
\bibitem{natsume}
Y. Natsume, Y. Watabe, T. Suzuki,
J. Phys. Soc. Jpn. {\bf 67} (1998) 3314.
\bibitem{freitas}
P. J. Freitas, R. R. P. Singh,
Phys. Rev. B {\bf 62} (2000) 14113.
\bibitem{schmidt}
K. P. Schmidt, C. Knetter, G. S. Uhrig,
Europhys. Lett. {\bf 56} (2001) 877.
\bibitem{takahashi}
M. Takahashi, T. Tohyama, S. Maekawa,
Phys. Rev. B {\bf 66} (2002) 125102.
\bibitem{tohyama}
T. Tohyama, H. Onodera, K. Tsutsui, S. Maekawa,
to appear in Phys. Rev. Lett.

\end{thebibliography}
\end{document}